\documentclass[envcountsame,runningheads]{llncs}

\usepackage{pdfpages}

\usepackage{amsmath,amssymb,amsfonts}
\usepackage{graphicx}

\usepackage{subfigure}
\usepackage{float}
\usepackage{bm}
\usepackage{multirow}
\usepackage{makecell}
\usepackage{color}
\usepackage{booktabs} 
\usepackage{caption} 
\usepackage{bbm}

\usepackage{tikz}
\usetikzlibrary{spy}

\usepackage{algorithm}
\usepackage{algorithmicx}
\usepackage{algpseudocode}

\newcommand{\y}		{\mathbf{y}}
\newcommand{\A}		{\mathbf{A}}

\newcommand{\W}	 {\mathbf{W}}

\newcommand{\0}	     {\mathbf{0}}

\newcommand{\I}	     {\mathbf{I}}

\newcommand{\R}	     {\mathsf{R}}

\newcommand*{\diag}{\mathrm{diag}}

\newcommand{\argmin}{\operatornamewithlimits{argmin}}

\newcommand{\alp}	 {\bm{\alpha}}

\def\R#1{(\ref{#1})}

\newcommand{\dquotes}[1]{``#1''}

\newcommand{\be}[1]{\begin{equation} #1 \end{equation}}
\newcommand{\bes}[1]{\begin{equation*} #1 \end{equation*}}
\newcommand{\ea}[1]{\begin{align} #1 \end{align}}

\newcommand*{\sgn}{\mathrm{sign}}

\usepackage{upgreek}
\newcommand{\upgreek}[1]{{
\renewcommand{\theta}{\uptheta}
\renewcommand{\rho}{\uprho}
\renewcommand{\nu}{\upnu}
\renewcommand{\psi}{\uppsi}
\renewcommand{\omega}{\upomega}
#1 }}

\newcommand*{\mb}[1]{{\mathbf{#1}}}
\newcommand*{\bbR}{\mathbb{R}}
\newcommand*{\cD}{\mathcal{D}}

\newcommand*{\cT}{\mathcal{T}}
\newcommand*{\rth}{\mathrm{th}}

\def\thetabf{\bm{\upgreek{\theta}}}
\def\rhobf{\bm{\upgreek{\rho}}}

\newcommand\blfootnote[1]{%
  \begingroup
  \renewcommand\thefootnote{}\footnote{#1}%
  \addtocounter{footnote}{-1}%
  \endgroup
}

\usepackage{wrapfig}

\makeatletter
\renewcommand{\ALG@name}{Algo.}
\makeatother

\newcommand{\specialcell}[2][c]{%
  \begin{tabular}[#1]{@{}c@{}}#2\end{tabular}}

\newcolumntype{C}[1]{>{\centering\let\newline\\\arraybackslash\hspace{0pt}}m{#1}}

\makeatletter
\renewcommand\section{\@startsection{section}{1}{\z@}%
                       {-8\p@ \@plus -4\p@ \@minus -4\p@}%
                       {6\p@ \@plus 4\p@ \@minus 4\p@}%
                       {\normalfont\large\bfseries\boldmath
                        \rightskip=\z@ \@plus 8em\pretolerance=10000 }}
\renewcommand\subsection{\@startsection{subsection}{2}{\z@}%
                       {-8\p@ \@plus -4\p@ \@minus -4\p@}%
                       {6\p@ \@plus 4\p@ \@minus 4\p@}%
                       {\normalfont\normalsize\bfseries\boldmath
                        \rightskip=\z@ \@plus 8em\pretolerance=10000 }}
\renewcommand\subsubsection{\@startsection{subsubsection}{3}{\z@}%
                       {-4\p@ \@plus -4\p@ \@minus -4\p@}%
                       {-1.5em \@plus -0.22em \@minus -0.1em}%
                       {\normalfont\normalsize\bfseries\boldmath}}

\renewcommand\subsubsection{\@startsection{subsubsection}{3}{\z@}%
                       {-8\p@ \@plus -4\p@ \@minus -4\p@}
                       {-0.5em \@plus -0.22em \@minus -0.1em}%
                       {\normalfont\normalsize\bfseries\boldmath}}
\makeatother

\begin{document}


\title{BCD-Net for Low-dose CT Reconstruction: Acceleration, Convergence, and Generalization}
\titlerunning{BCD-Net for Low-dose CT Reconstruction}
\author{
Il Yong Chun$^{\ast,}$\inst{1} \and
Xuehang Zheng$^{\ast,}$\inst{2} \and 
Yong Long\inst{2} \and
Jeffrey A. Fessler\inst{1}
}


\authorrunning{I.Y.~Chun and X.~Zheng et al.}
\institute{
Department of Electrical Engineering and Computer Science, \\
University of Michigan, MI, USA \\ \and 
University of Michigan - Shanghai Jiao Tong University Joint Institute, \\
Shanghai Jiao Tong University, Shanghai, China 
}

\maketitle              
%

\begin{abstract}
Obtaining accurate and reliable images from low-dose computed tomography (CT) is challenging.
Regression convolutional neural network (CNN) models that are learned from training data are increasingly gaining attention in low-dose CT reconstruction.
This paper modifies the architecture of an iterative regression CNN, \emph{BCD-Net}, for fast, stable, and accurate low-dose CT reconstruction, and presents the convergence property of the modified BCD-Net.
Numerical results with phantom data show that applying faster numerical solvers to model-based image reconstruction (MBIR) modules of BCD-Net leads to faster and more accurate BCD-Net; 
BCD-Net significantly improves the reconstruction accuracy, compared to the state-of-the-art MBIR method using learned transforms; 
BCD-Net achieves better image quality, compared to a state-of-the-art iterative NN architecture, ADMM-Net.
Numerical results with clinical data show that BCD-Net generalizes significantly better than a state-of-the-art deep (non-iterative) regression NN, FBPConvNet, that lacks MBIR modules.
\blfootnote{The authors indicated by asterisks ($^\ast$) equally contributed to this work.}
\blfootnote{Corresponding author: Yong Long (email: \email{yong.long@sjtu.edu.cn}).}
\blfootnote{This paper has supplementary document. The prefix \dquotes{S} indicates the numbers in figure and section in the supplementary document.}
\end{abstract}

\section{Introduction}     

Low-dose computed tomography (CT) reconstruction requires careful regularization design to control noise while preserving crucial image features.
Traditional regularizers have been based on mathematical models like total variation, whereas newer methods are based on models that are learned from training data, especially regression neural network (NN) models.
Deep convolutional NN (CNN) methods in an early stage map low- to high-quality images: specifically, they \dquotes{denoise} the artifacts in the low-quality images obtained by applying some basic solvers to raw data or measurements.
However, the greater mapping capability (i.e., higher the NN complexity) can increase the overfitting risks \cite{Zheng&etal:19TCI}.
There exist several ways to prevent NNs from overfitting, 
e.g., increasing the dataset size, reducing the neural network complexity, and dropout.
However, in solving large-scale inverse problems in imaging, 
the first scheme is limited in training CNNs from large-scale images; 
the second scheme does not effectively remove complicated noise features; 
and the third scheme has limited benefits when applied to convolutional layers.

An alternative way to regulate overfitting of regression CNNs in inverse imaging problems is 
combining them with model-based image reconstruction (MBIR) that considers imaging physics or image formation models, and noise statistics in measurements.
BCD-Net \cite{Chun&Fessler:18IVMSP} is an iterative regression CNN that generalizes a block coordinate descent (BCD) MBIR method using learned convolutional regularizers \cite{Chun&Fessler:18arXiv}.
Each layer (or iteration) of BCD-Net consists of image denoising and MBIR modules.
In particular, the denoising modules use layer-wise regression CNNs to effectively remove layer-dependent noise features.
Many existing works can be viewed as a special case of BCD-Net.
For example, 
RED \cite{Romano&Elad&Milanfar:17SJIS} and MoDL \cite{Aggarwal&Mani&Jacobs:18TMI} are special cases of BCD-Net, 
because they use identical image denoising modules across layers or only consider quadratic data-fidelity terms (e.g., the first term in \R{sys:recon}) in their MBIR modules.

This paper modifies BCD-Net that uses convolutional autoencoders in its denoising modules \cite{Chun&Fessler:18IVMSP}, and applies the modified BCD-Net to low-dose CT reconstruction.
First, for fast CT reconstruction, we apply the Accelerated Proximal Gradient method using a Majorizer (APG-M), e.g., FISTA \cite{Beck&Teboulle:09SIAM}, to MBIR modules using the statistical CT data-fidelity term.
Second, this paper provides the sequence convergence guarantee of BCD-Net when applied to low-dose CT reconstruction.
Third, it investigates the generalization capability of BCD-Net for low-dose CT reconstruction, compared to a state-of-the-art deep (non-iterative) regression NN, FBPConvNet \cite{Jin&etal:17TIP}.
Numerical results with the extended cardiac-torso (XCAT) phantom show that applying faster numerical solvers (e.g., APG-M) to MBIR modules leads to faster and more accurate BCD-Net;
regardless of numerical solvers of MBIR modules, BCD-Net significantly improves the reconstruction accuracy, compared to the state-of-the-art MBIR method using learned transforms \cite{Zheng&etal:19TCI}; given identical denoising CNN architectures, BCD-Net achieves better image quality, compared to a state-of-the-art iterative NN architecture, ADMM-Net \cite{Yang&etal:16NIPS}.
Numerical results with clinical data show that BCD-Net generalizes significantly better than FBPConvNet \cite{Jin&etal:17TIP} that lacks MBIR modules.

\section{BCD-Net for Low-Dose CT Reconstruction}

\subsection{Architecture}

\begin{wrapfigure}{r}{0pt}
	\centering
	\begin{minipage}{0.52\linewidth}
		\vspace{-2.75pc}
		\begin{algorithm}[H] 
			\captionof{algorithm}{BCD-Net for CT reconstruction}
			\label{alg:bcd-net}
			\begin{algorithmic}[0] 
				\Require $\{\cD_{\thetabf^{(l)}} : \forall l \}, \mb{x}^{(0)}, \mb{y}, \mb{A}, \mb{W}, \beta$
				\For {$l = 0, \ldots, L-1$}  
				\State \emph{Denoising}: $\displaystyle \mb{z}^{(l+1)} = \cD_{\thetabf^{(l+1)}} (\mb{x}^{(l)})$
				\State \emph{MBIR}: $\displaystyle \mb{x}^{(l+1)} = \argmin_{\mb{x} \geq \mb{0}}  F(\mb{x}; \mb{y}, \mb{z}^{(l+1)})$$^\dagger$
				\EndFor
				\State \hspace{-1.0em}$^\dagger$$F(\mb{x}; \mb{y}, \mb{z}^{(l+1)})$ is defined in \R{sys:recon}. 
			\end{algorithmic}
		\end{algorithm}
		\vspace{-2.75pc}
	\end{minipage}
\end{wrapfigure}

This section modifies the architecture of BCD-Net in \cite{Chun&Fessler:18IVMSP} for CT reconstruction.
For the image denoising modules, we use layer-wise autoencoding CNNs that apply exponential function to trainable thresholding parameters.
(The trainable thresholding parameters replace the bias terms, since biases can differ greatly for different objects in imaging problems.)
The layer-wise denoising CNNs are particularly useful to remove layer-dependent artifacts in reconstructed images at the previous layers, without greatly increasing their parameter dimensions.
In low-dose CT reconstruction, for example, the CNNs at the early and later layers remove streak artifacts and Gaussian-like noise, respectively.
MBIR modules aim to regularize overfitting artifacts by combining information drawn from the data-fidelity term and output of denoising modules.
Different from the image denoising and single-coil magnetic resonance imaging applications in \cite{Chun&Fessler:18IVMSP}, the MBIR modules of CT reconstruction BCD-Net involve iterative solvers. 
For fast CT reconstruction in particular, we apply a fast numerical solver, APG-M, to the MBIR modules.
Algo.~\ref{alg:bcd-net} shows the architecture of the modified BCD-Net for CT reconstruction.

\subsubsection{Image Denoising Module.}

For the $l\rth$ layer image denoising module, we use a convolutional autoencoder in the following form:
\be{
\label{eq:autoencoder}
\cD_{\thetabf^{(l+1)}} (\cdot) =  \frac{1}{R} \sum_{k=1}^{K} \mb{d}_k^{(l+1)}  \circledast \cT_{\exp(\alpha_k^{(l+1)})} ( \mb{e}_k^{(l+1)} \circledast (\cdot) ),
}
where $\thetabf^{(l+1)} :=  \{ \mb{d}_k^{(l+1)}, \mb{e}_k^{(l+1)}, \alpha_k^{(l+1)} : k \!=\! 1,\ldots,K, l \!=\! 1,\ldots,L \}$ is a parameter set of the $l\rth$ convolutional autoencoder,
$\mb{d}_k^{(l+1)}  \in \bbR^R$, $\mb{e}_k^{(l+1)} \in \bbR^R$, and $\exp(\alpha_k^{(l+1)})$ are the $k\rth$ decoding and encoding filters, and thresholding value at the $l\rth$ layer, respectively,
the convolution operator $\circledast$ uses the circulant boundary condition without the filter flip, 
$\cT_{a}(\cdot)$ is the soft-thresholding operator with the thresholding parameter $a$,
$R$ and $K$ are the size and number of the filters, respectively, and $L$ is the number of layers in BCD-Net.
Different from the original convolutional autoencoder in \cite{Chun&Fessler:18IVMSP}, we included the exponential function $\exp(\cdot)$ to prevent the thresholding parameters $\{ \alpha_k \}$ from becoming zero during training \cite{Chun&etal:18arXiv:momnet}.
The factor $1/R$ comes from the relation between convolution-perspective and patch-based trainings \cite{Chun&etal:18arXiv:momnet}.
By applying the trained convolutional autoencoder in \R{eq:autoencoder} to the $l\rth$ layer input $\mb{x}^{(l)}$ (i.e., reconstructed image at the $(l-1)\rth$ layer), 
we obtain the \dquotes{denoised} image $\mb{z}^{(l+1)} = \cD_{\thetabf^{(l+1)}} (\mb{x}^{(l)})$. 
We next feed $\mb{z}^{(l+1)}$ into the $l\rth$ layer MBIR module.

\subsubsection{MBIR Module.}

The $l\rth$ layer MBIR module uses the $l\rth$ layer denoised image $\mb{z}^{(l+1)}$, and reconstructs an image $\mb{x} \in \bbR^{N}$ from post-log measurement $\mb{y} \in \bbR^{M}$ by solving the following statistical MBIR problem: 
\be{
\label{sys:recon}	
\mb{x}^{(l+1)} = \argmin_{\mb{x} \succeq \0} F(\mb{x}; \mb{y}, \mb{z}^{(l+1)}) :=\frac{1}{2}\|\mb{y} - \mb{A} \mb{x}\|^2_{\mb{W}}  + \frac{\beta}{2} \|\mb{x} - \mb{z}^{(l+1)}\|^2_2,
\tag{P1}
}
where $\mb{A} \in  \mathbb{R}^{  M  \times N}$ is a CT scan system matrix, 
$\mb{W}  \in \bbR^{M \times M}$ is a diagonal weighting matrix with elements $\{ W_{m,m} \!=\! \rho_m^2 / ( \rho_m + \sigma^2 ) \!:\! \forall m \}$ 
based on a Poisson-Gaussian model for the pre-log measurements $\rhobf \in \bbR^M$ with electronic readout noise variance $\sigma^2$ \cite{Zheng&etal:19TCI}, 
and $\beta>0$ is a regularization parameter.
To rapidly solve \R{sys:recon}, we apply APG-M, a generalized version of APG (e.g., FISTA \cite{Beck&Teboulle:09SIAM}) that uses $M$-Lipschitz continuous gradients \cite{Chun&Fessler:18arXiv}. Initialized with $\mb{v}^{(0)} = \bar{\mb{x}}^{(0)} = \mb{x}^{(l)}$ and $t_0 = 1$, the APG-M updates are
\ea{
\label{eq:apg-m:x}
\bar{\mb{x}}^{(j+1)} &= \Big[\mb{v}^{(j)} + \mb{M}^{-1} \big( \A^T \W (\y - \A \mb{v}^{(j)}) - \beta (\mb{v}^{(j)} - \mathbf{\mb{z}}^{(l+1)}) \big) \Big]_+, 
\\
\label{eq:apg-m:v}
\mb{v}^{(j+1)} &= \bar{\mb{x}}^{(j+1)}+\frac{t_{j} -1 }{t_{j+1}} ( \bar{\mb{x}}^{(j+1)} -  \bar{\mb{x}}^{(j)}), \quad \mbox{where}~t_{j+1} = (1+\sqrt{1+4t_j^2}) / 2, 
}	
for $ j = 0,\ldots,J-1$,
where the operator $[\cdot]_+$ is the proximal operator obtained by considering the non-negativity constraint in \R{sys:recon} and clips the negative values, and $J$ is the number of APG-M iterations.
We design the diagonal majorizer $\mb{M} \in \bbR^{N \times N}$ in \R{eq:apg-m:x} as follows \cite{Chun&Fessler:18arXiv}:
$\mb{M} = \diag( \mb{A}^{T}\mb{W}\mb{A} \mathbf{1} ) + \beta \I  \succeq  \nabla^2 F(\mb{x}; \mb{y}, \mb{z}^{(l+1)}) = \mb{A}^{T}\mb{W}\mb{A} + \beta \I$, 
where $\diag(\cdot)$ converts a vector into a diagonal matrix.
The $l\rth$ layer reconstructed image $\mb{x}^{(l+1)}$ is given by the $J\rth$ APG-M update, i.e., $\mb{x}^{(l+1)} = \bar{\mb{x}}^{(J)}$, and fed into the next BCD-Net layer as an input.

\subsection{Training BCD-Net} \label{sec:training}

\begin{wrapfigure}{r}{0pt}
\centering
\begin{minipage}{0.57\linewidth}
\vspace{-2.75pc}
\begin{algorithm}[H]  
	\caption{Training BCD-Net for CT recon.} 
	\label{alg:training}
	\begin{algorithmic}[0]
		\Require 
		$\{ \mb{x}_i, \mb{x}_i^{(0)}, \mb{y}_i, \mb{A}, \mb{W}_i, \beta : \forall i \}$
		\For {$l = 0, \ldots, L-1$}
		\State \emph{Train} $\thetabf^{(l +1)}$: Solve \R{sys:training} using $\{ \mb{x}_i, \mb{x}_i^{(l)} : \forall i \}$
		\For{$i = 1,\ldots, I $ } 
		
		\State \emph{Denoising}:~$\mb{z}_i^{(l+1)} \!= \cD_{\thetabf^{(l+1)}}( \mb{x}_i^{(l)} )$
		\State \emph{MBIR}:~$\displaystyle \mb{x}_i^{(l+1)} \!= \argmin_{\mb{x} \geq \mb{0}}  F_{i} (\mb{x}; \mb{y}_i, \mb{z}_i^{(l+1)})$$^\dagger$ 

		\EndFor
		\EndFor
		\State \hspace{-1.0em}$^\dagger$$F(\mb{x}; \mb{y}, \mb{z}^{(l+1)})$ is defined in \R{sys:recon}. 
	\end{algorithmic}
\end{algorithm}
\vspace{-2.75pc}
\end{minipage}
\end{wrapfigure}

This section proposes a BCD-Net training framework for CT reconstruction, based on the image denoising and MBIR modules defined in the previous section. 
The training process requires $I$ high-quality training images, $\{ \mb{x}_i : i  = 1,\ldots,I \}$, 
and $I$ training measurements simulated via CT physics, $\{ \mb{y}_i : i  = 1,\ldots,I \}$.
Algo.~\ref{alg:training} summarizes the training framework.

At the $l\rth$ layer, we optimize the parameters $\thetabf^{(l+1)}$ of the $l\rth$ convolutional autoencoder in \R{eq:autoencoder} from $I$ training pairs $(  \mb{x}_i, \mb{x}_i^{(l)}  )$, where $\mb{x}_i^{(l)}$ is the $i\rth$ reconstructed training image at the $(l-1)\rth$ layer.
Our patch-based training loss function at the $l\rth$ layer is 
\be{
\label{sys:training}
\thetabf^{(l+1)} = \argmin_{\{ \mb{D}, \alp, \mb{E} \}} \frac{1}{R \tilde{P}}\| \widetilde{\mb{X}} -  \mb{D}  \cT_{\exp(\alp)}  ( \mb{E}^T \widetilde{\mb{X}}^{(l)} ) \|_F^2,
\tag{P2}
}
where encoding and decoding filter matrices $\mb{D} \in \mathbb{R}^{R\times K}$ and $\mb{E} \in \mathbb{R}^{R\times K}$ 
are formed by grouping $K$ filters as $\mb{D} : = [\mb{d}_1, \ldots, \mb{d}_K]$ and $\mb{E} : = [\mb{e}_1, \ldots, \mb{e}_K]$, 
respectively, $\alp \in \mathbb{R}^{K}$ is a vector containing $K$ thresholding values,
$\tilde{P}$ is the number of patches extracted from all training images,
and $\widetilde{\mb{X}} \in \bbR^{R \times \tilde{P}}$ and $\widetilde{\mb{X}}^{(l)} \in \bbR^{R \times \tilde{P}}$ 
are paired training matrices whose columns are vectorized patches extracted from $\{ \mb{x}_i : \forall i  \}$ and $\{ \mb{x}_i^{(l)} : \forall i \}$, respectively.
The soft thresholding operator $\cT_{\mb{a}}(\mb{u}) : \mathbb{R}^{K} \rightarrow \mathbb{R}^{K}$ is defined as follows: $( \cT_{\mb{a}} (\mb{u}) )_{k}$ equals to $u_{k} - a_{k} \sgn (u_{k})$ for $ | u_{k} | > a_{k}$, and $0$ otherwise, $\forall k$.
We optimize \R{sys:training} via a mini-batch stochastic gradient method.

\subsection{Convergence Analysis}

There exist two key challenges in understanding the convergence behavior of BCD-Net in Algo.~\ref{alg:bcd-net}:
\textit{1)} (general) denoising NNs $\{ \cD_{\thetabf^{(l+1)}} \}$ change across layers;  
\textit{2)} even if they are identical across layers, they are not necessarily nonexpansive operators \cite{Rockafellar:76SIAMCO} in practice. 
To moderate these issues, we introduce a new definition:
\begin{definition}
[Asymptotically nonexpansive paired operators \mbox{\cite{Chun&etal:18arXiv:momnet}}] 
\label{d:pair-map} 
\\
Paired operators $\{ \cD_{\thetabf^{(l)}}, \cD_{\thetabf^{(l+1)}} \}$ are asymptotically nonexpansive if there exist a summable sequence $\{ \epsilon^{(l+1)} \in [0,\infty) : \sum_{l=0}^\infty \epsilon^{(l+1)} < \infty \}$ such that
\bes{
\left\| \cD_{\thetabf^{(l+1)}} (\mb{u}) - \cD_{\thetabf^{(l)}} (\mb{v}) \right\|_2^2 \leq \| \mb{u} - \mb{v} \|_2^2 + \epsilon^{(l+1)}, \qquad \mbox{$\forall \mb{u},\mb{v}$ and $\forall l$}.
}
\end{definition}

Based on Definition~\ref{d:pair-map}, we obtain the following convergence result for Algo.~\ref{alg:bcd-net}:

\begin{theorem}
[Sequence convergence] 
\label{t:seq:convg}
Assume that paired denoising neural networks $\{ \cD_{\thetabf^{(l)}}, \cD_{\thetabf^{(l+1)}} \}$ are asymptotically nonexpansive with the summable sequence $\{ \epsilon^{(l+1)} \in [0,\infty) : \sum_{l=1}^\infty \epsilon^{(l+1)} < \infty \}$ and $\mb{A}^T \mb{W} \mb{A} \succ 0$. 
Then the sequence $\{ \mb{x}^{(l+1)} : l \geq 0 \}$ generated by Algo.~\ref{alg:bcd-net} (disregarding the non-negativity constraints in the MBIR optimization problems \R{sys:recon}) is convergent.
\end{theorem}

Theorem~\ref{t:seq:convg} implies that if denoising neural networks $\{ \cD_{\thetabf^{(l)}} : l \geq 1 \}$ converge to a nonexpnasive one, BCD-Net guarantees the sequence convergence.
Fig.~S.1 shows the convergence behaviors of $\{ \cD_{\thetabf^{(l+1)}} \}$ and their Lipschitz constants.

\subsection{Computational complexity}

The computational cost of the proposed BCD-Net is $O((MJ+RK)NL)$. Since $MJ \gg RK$, the computational complexity of BCD-Net is dominated by forward and back projections performed in the MBIR modules. To reduce the $MJ$ factor, one can investigate faster optimization methods (e.g., proximal optimized gradient method (POGM) \cite{taylor:17:ewc-composite}) with ordered subsets. 
Applying these techniques can reduce the $MJ$ factor to $(M/G) J'$, where $G$ is the number of subsets and the number of POGM iterations $J' < J$ (e.g., $J' = (1/\sqrt{2}) J$) due to faster convergence rates of POGM over APG.

\section{Experimental Results and Discussion}

\subsection{Experimental Setup} \label{sec:exp:imag}

\subsubsection{Imaging.}   

For XCAT phantom images \cite{Segars&etal:08MP} and reconstructed clinical images in \cite{Zheng&etal:19TCI}, we simulated sinograms of size $888 \!\times\!984$ (detectors$\,\times\,$projection views) with GE LightSpeed fan-beam geometry corresponding to a monoenergetic source with $\rho_0 \!=\! 10^4$ incident photons per ray and electronic noise variance $\sigma^2 \!=\! 5^2$ \cite{Zheng&etal:19TCI} (while avoiding inverse crimes). 
We reconstructed $420 \!\times\! 420$ images with pixel-size $\Delta_x \!=\! \Delta_y \!=\! 0.9766$ mm.
For the clinical data collected from the GE scanner using the CT geometry above, and tube voltage $120$ kVp and current $160$ mA, 
we reconstructed a $716 \!\times\! 716$ image (shown in the third row of Fig.~\ref{fig:recon:ge}) with $\Delta_x \!=\! \Delta_y \!=\! 0.9777$ mm.

\subsubsection{Training BCD-Net, ADMM-Net, and FBPConvNet.}

Based on the proposed framework in Section~\ref{sec:training}, 
we trained $100$-layer BCD-Nets with the two parameter sets, $\{ K \!=\! R \!=\! 8^2 \}$ and $\{ K \! = \! 10^2, R \!=\! 8^2 \}$,
and the regularization parameter $\beta \!=\! 4 \!\times\! 10^6$.
In particular, we solved \R{sys:training} with Adam \cite{Kingma&Ba:15ICLR} and $\tilde{P} \!\approx\! 1.7 \!\times\! 10^6$ training patches that were extracted from ten training images of the XCAT phantom \cite{Segars&etal:08MP}. 
We used the mini-batch size $512$, $200$ epochs, 
initial learning rates $10^{-3}$, and $10^{-2}$ for $\{ \mb{D}^{(l)}, \mb{E}^{(l)} : \forall l \}$ and $\{ \alp^{(l)} : \forall l \}$, 
and random i.i.d. Gaussian filter initialization.
We decayed the learning rates by a factor of $0.9$ every $10$ epochs.
We trained a $100$-layer ADMM-Net that uses the layer-wise denoising NNs \R{eq:autoencoder} with $ K \!=\! R \!=\! 8^2$,
with the identical training setup above.
We chose the ADMM penalty parameter as $1 \!\times\!10^6$, by matching the spatial resolution in the heart region of test sample $\#1$ to that reconstructed by BCD-Net. 
We trained FBPConvNet with $500$ 2D XCAT phantom images and the similar parameters suggested in \cite{Jin&etal:17TIP}.

\subsubsection{Image Reconstruction.}

We compared trained BCD-Nets with the conventional MBIR method using an edge-preserving (EP) regularizer, 
the state-of-the-art MBIR method using $\ell_2$ prior with a learned square transform \cite{Zheng&etal:19TCI}, 
a state-of-the-art iterative NN architecture, ADMM-Net \cite{Yang&etal:16NIPS} (i.e., plug-and-play ADMM \cite{chan:17:pap} using denoising NNs),
and/or a state-of-the-art (non-iterative) deep regression NN, FBPConvNet \cite{Jin&etal:17TIP}.
For the first two MBIR methods, we finely tuned their parameters to give the lowest root-mean-square-error (RMSE) values \cite{Chun&Fessler:18arXiv}.
(See their parameter details in Section~S.2).
We tested the aforementioned methods to two sets of three representative chest CT images that are selected from the XCAT phantom and clinical data provided by GE.
(Note that the testing phantom images are sufficiently different from training phantom images; specifically, they are $\approx$2cm away from training images.)
We quantitatively evaluated the quality of phantom reconstructions by RMSE (in Hounsfield units, HU) in a region of interest \cite{Zheng&etal:19TCI}.

\subsection{Results and Discussion}

\subsubsection{Convergence of BCD-Net with Different MBIR Modules.}

\begin{wrapfigure}{r}{0pt}
\centering
\hspace{0.2pc}
\begin{minipage}{0.44\linewidth}
\vspace{-2.75pc}
\begin{figure}[H] 
	\centering
	\includegraphics[trim=0.5em 0.25em 3.2em 2.4em,clip, width=5.5cm]{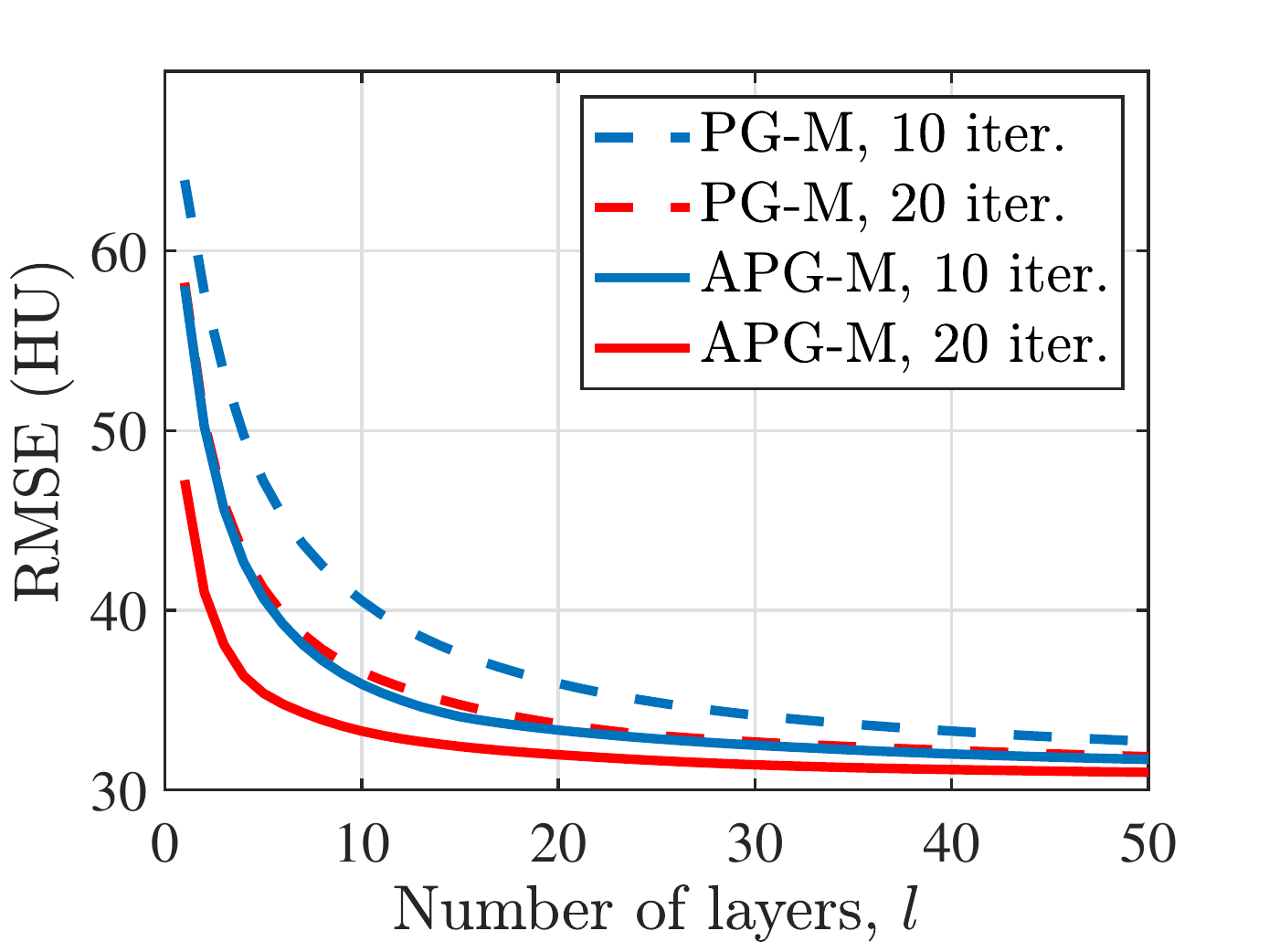}\vspace{-0.75em}
	\caption{RMSE convergence of BCD-Nets using different MBIR modules for low-dose CT reconstruction (for the first testing image in Table~\ref{tab:rmse:xcat}).}
\label{fig:convg}	 
\end{figure}
\vspace{-2.75pc}
\end{minipage}
\end{wrapfigure}

Applying faster iterative solvers to MBIR modules leads to faster and more accurate BCD-Net.
This assertion is supported by comparing the APG-M and PG-M results in Fig.~\ref{fig:convg} (given the identical iteration numbers), 
and noting that APG-M is faster than PG-M (i.e., APG-M using no \dquotes{momentum}, $\bar{\mb{x}}^{(j+1)} -  \bar{\mb{x}}^{(j)}$ in \R{eq:apg-m:v}).
In addition, Fig.~\ref{fig:convg} shows that increasing the number of iterations in numerical MBIR solvers leads more accurate BCD-Net, 
given the identical numbers of BCD-Net layers.
This implies that numerical MBIR solvers using insufficient number of iterations 
do not fully extract \dquotes{desired} information from CT data-fidelity (i.e., the first term in \R{sys:recon}).
The importance of using rapidly converging MBIR solvers is underestimated in existing literature: existing literature often considers some applications that have practical and closed-form MBIR solution \cite{Chun&Fessler:18IVMSP}.

\subsubsection{Reconstruction Quality Comparisons.}

\begin{figure*}[!t]
	\centering
	\small\addtolength{\tabcolsep}{-1pt}
	\begin{tabular}{ccc}
		\small{EP} & \small{FBPConvNet \mbox{\cite{Jin&etal:17TIP}}} & \small{\textbf{BCD-Net} ($K \!=\! R \!=\! 8^2$)} 
		\\
		
		\includegraphics[bb={18.5mm 17mm 98.5mm 97mm},clip,width=40mm,height=40mm]{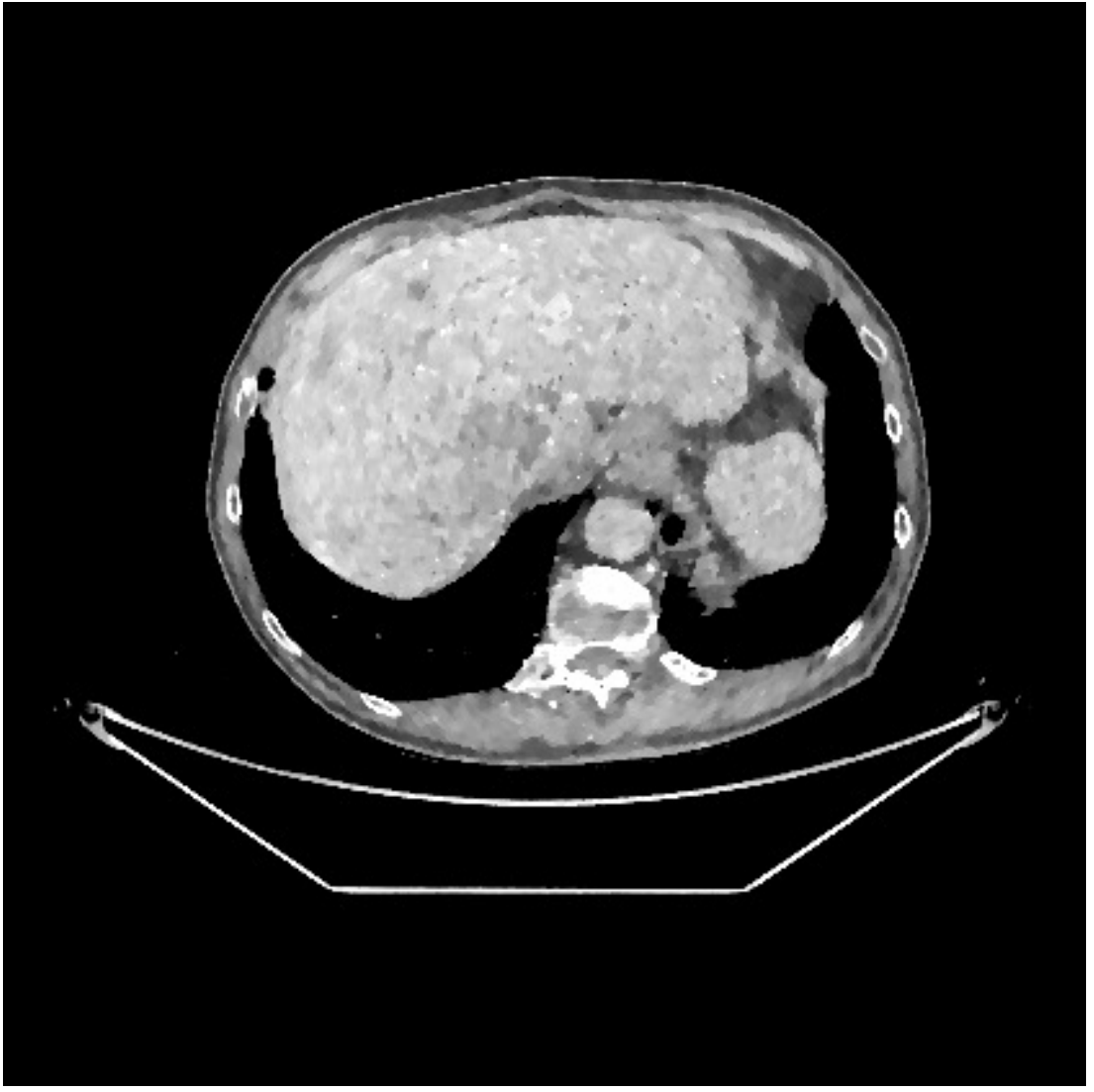}
		&
		\includegraphics[bb={18.5mm 17mm 98.5mm 97mm},clip,width=40mm,height=40mm]{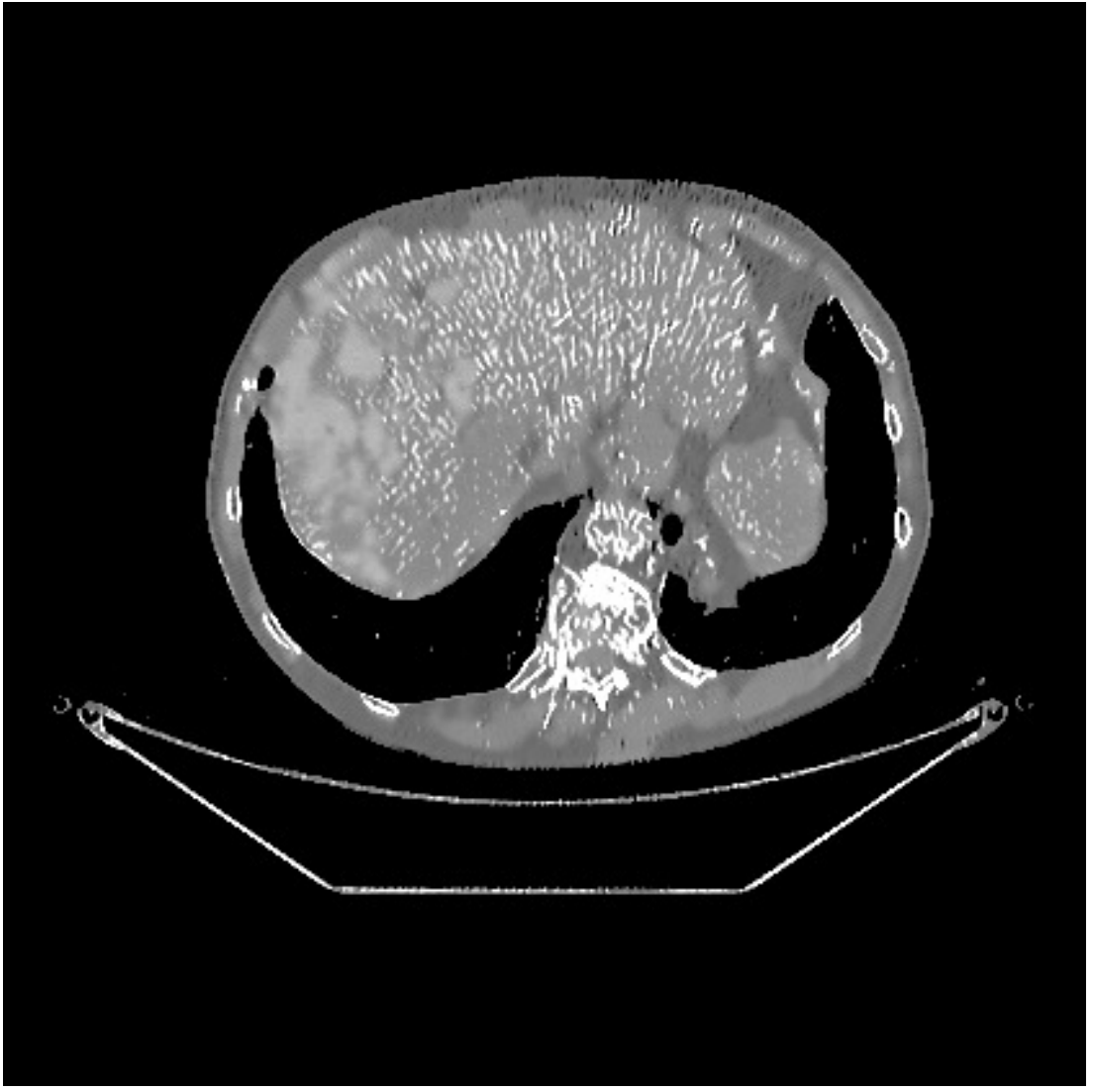}
		&
		\includegraphics[bb={18.5mm 17mm 98.5mm 97mm},clip,width=40mm,height=40mm]{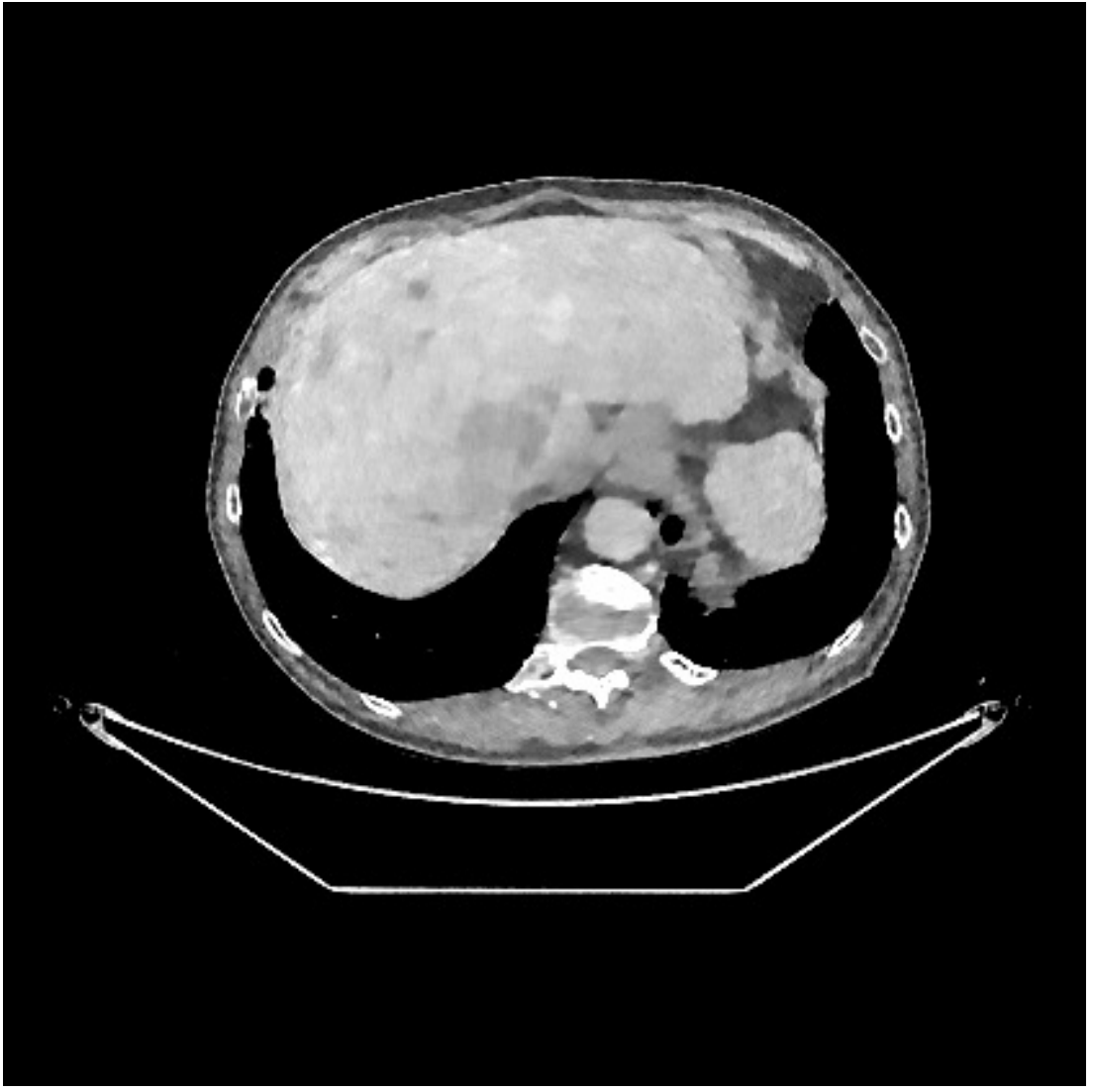}
		\\
				
		\includegraphics[bb={18.5mm 17mm 98.5mm 97mm},clip,width=40mm,height=40mm]{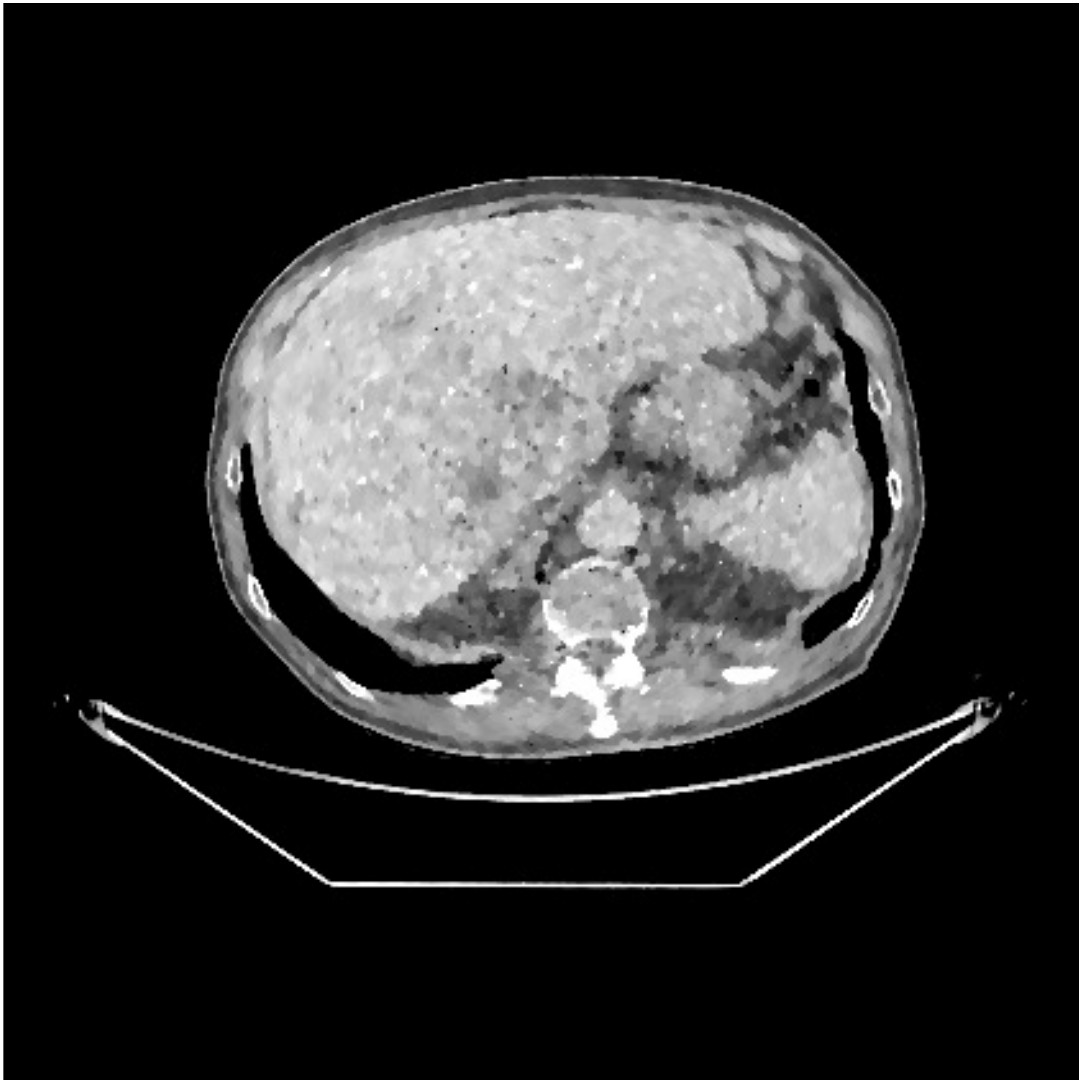}
		&
		\includegraphics[bb={18.5mm 17mm 98.5mm 97mm},clip,width=40mm,height=40mm]{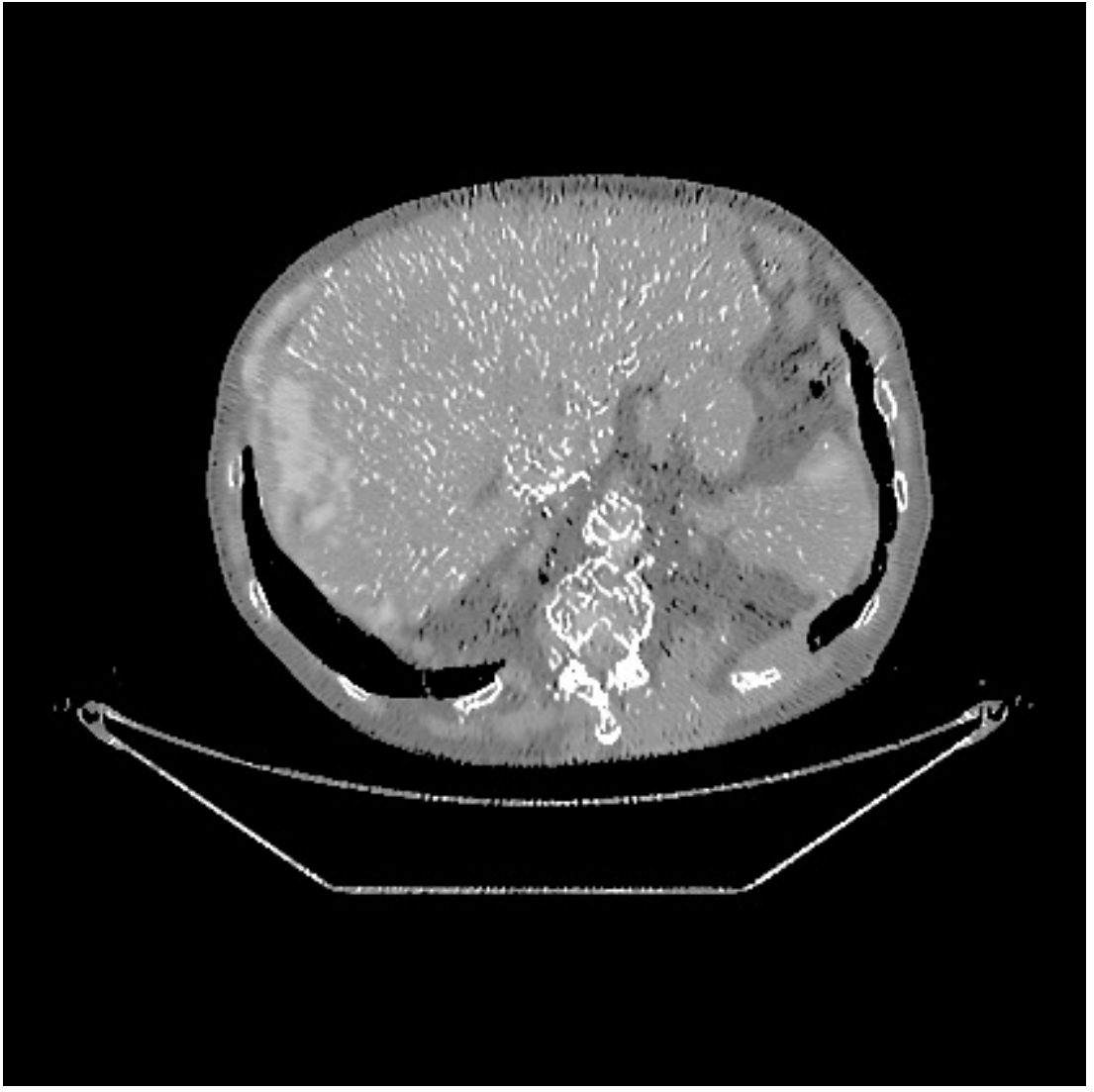}
		&
		\includegraphics[bb={18.5mm 17mm 98.5mm 97mm},clip,width=40mm,height=40mm]{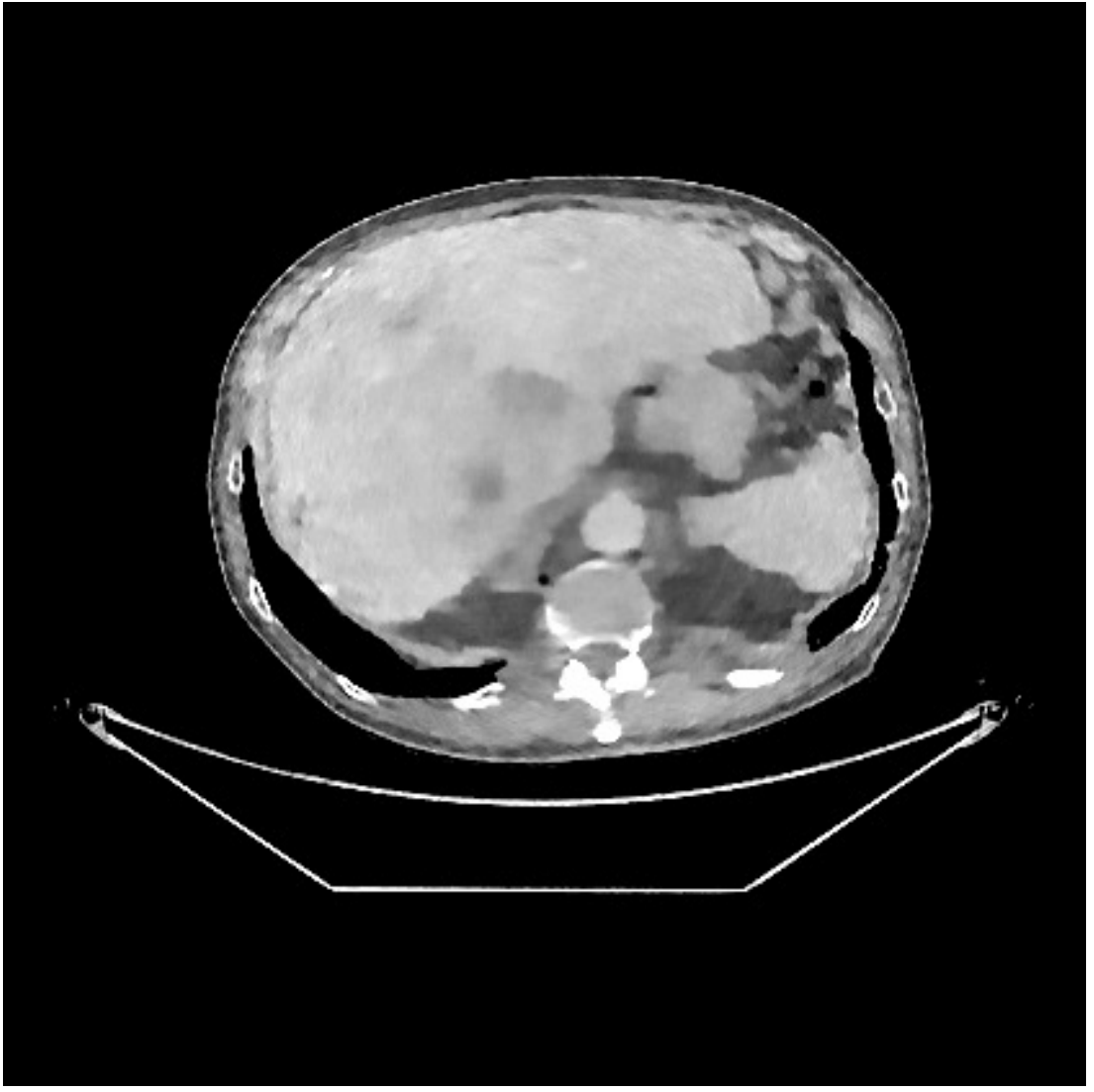}
		\\
		
		\includegraphics[bb={18mm 18mm 118mm 118mm},clip,width=40mm,height=40mm]{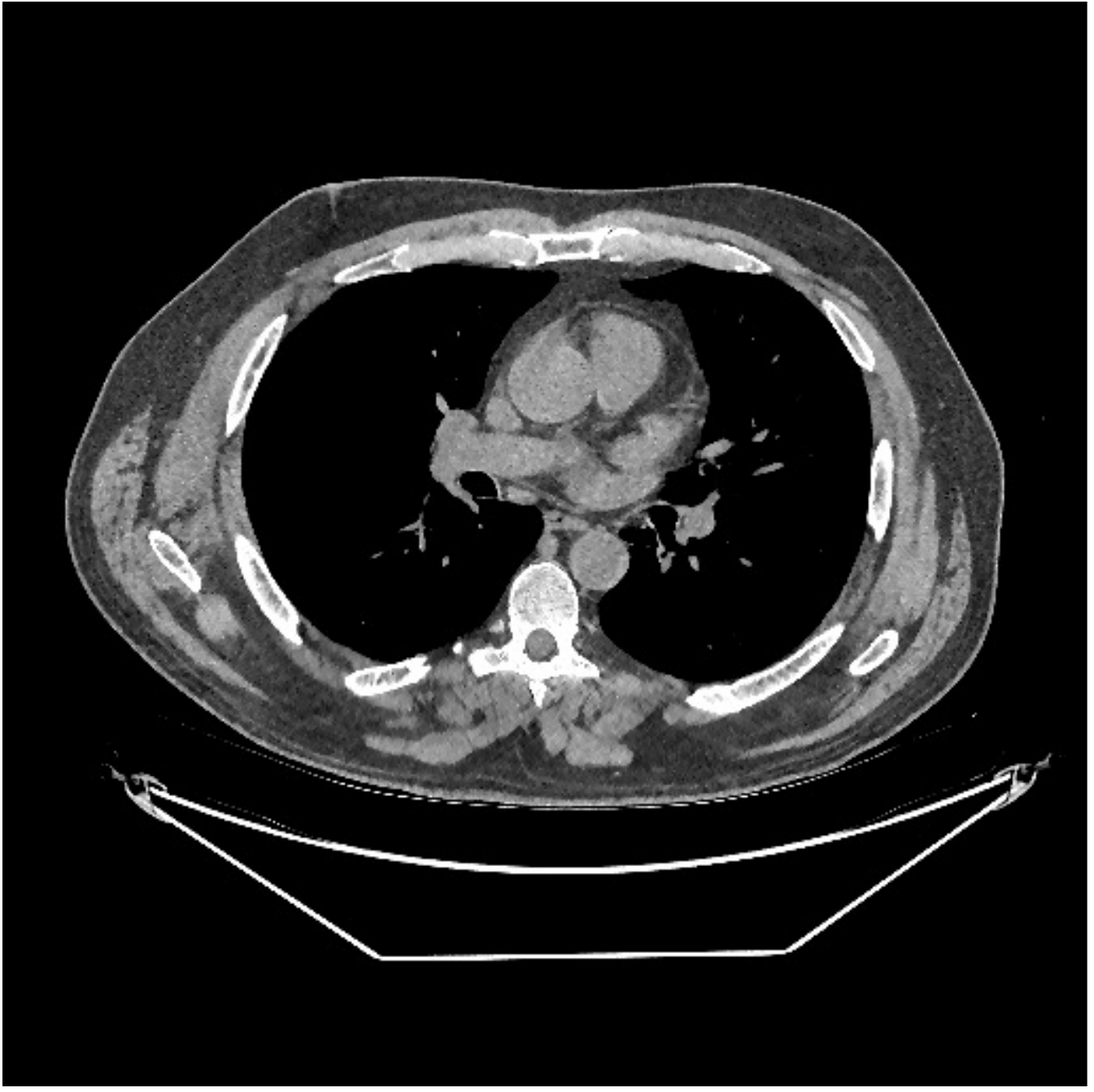}
		&
		\includegraphics[bb={18mm 18mm 118mm 118mm},clip,width=40mm,height=40mm]{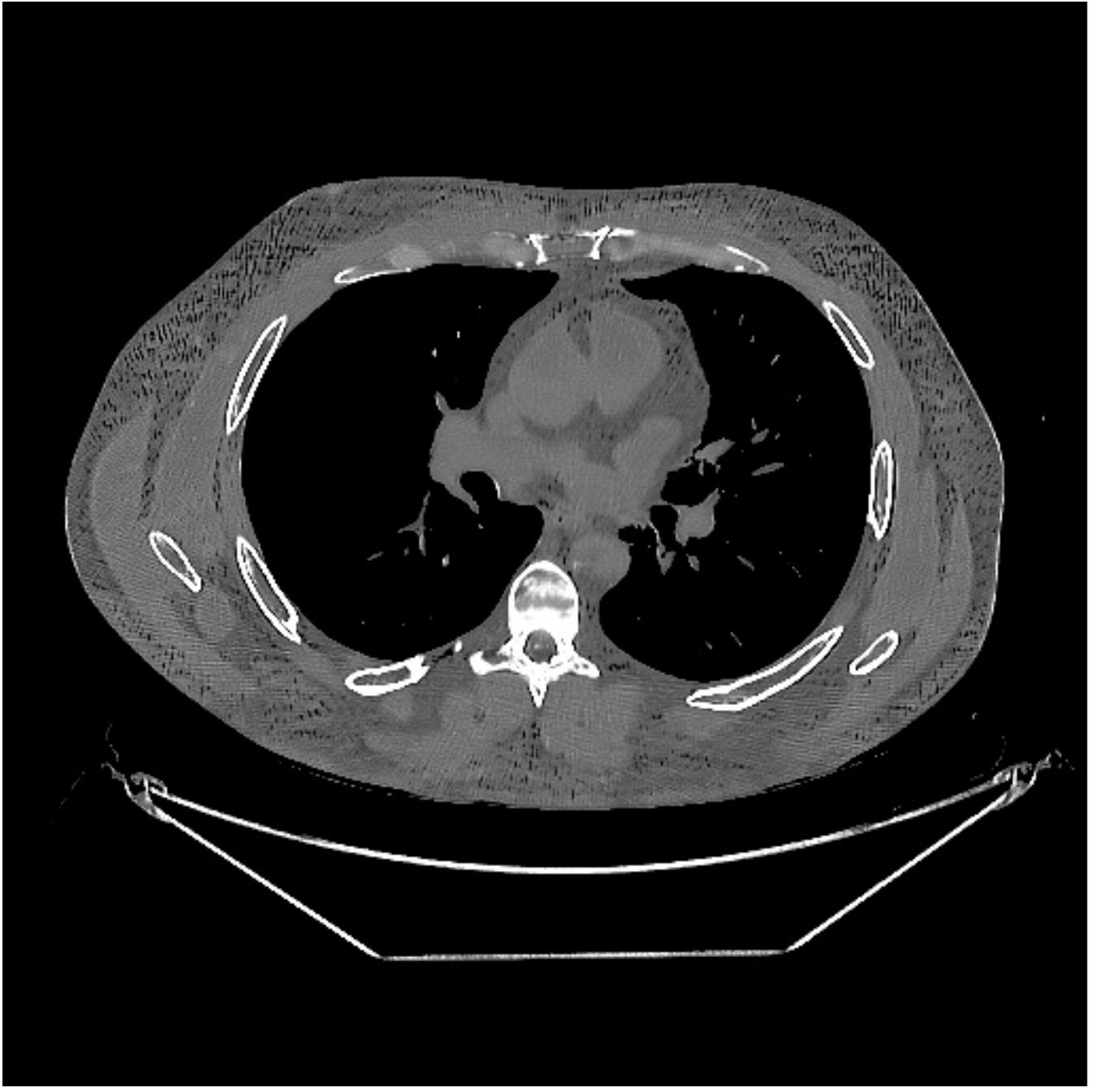}
		&
		\includegraphics[bb={18mm 18mm 118mm 118mm},clip,width=40mm,height=40mm]{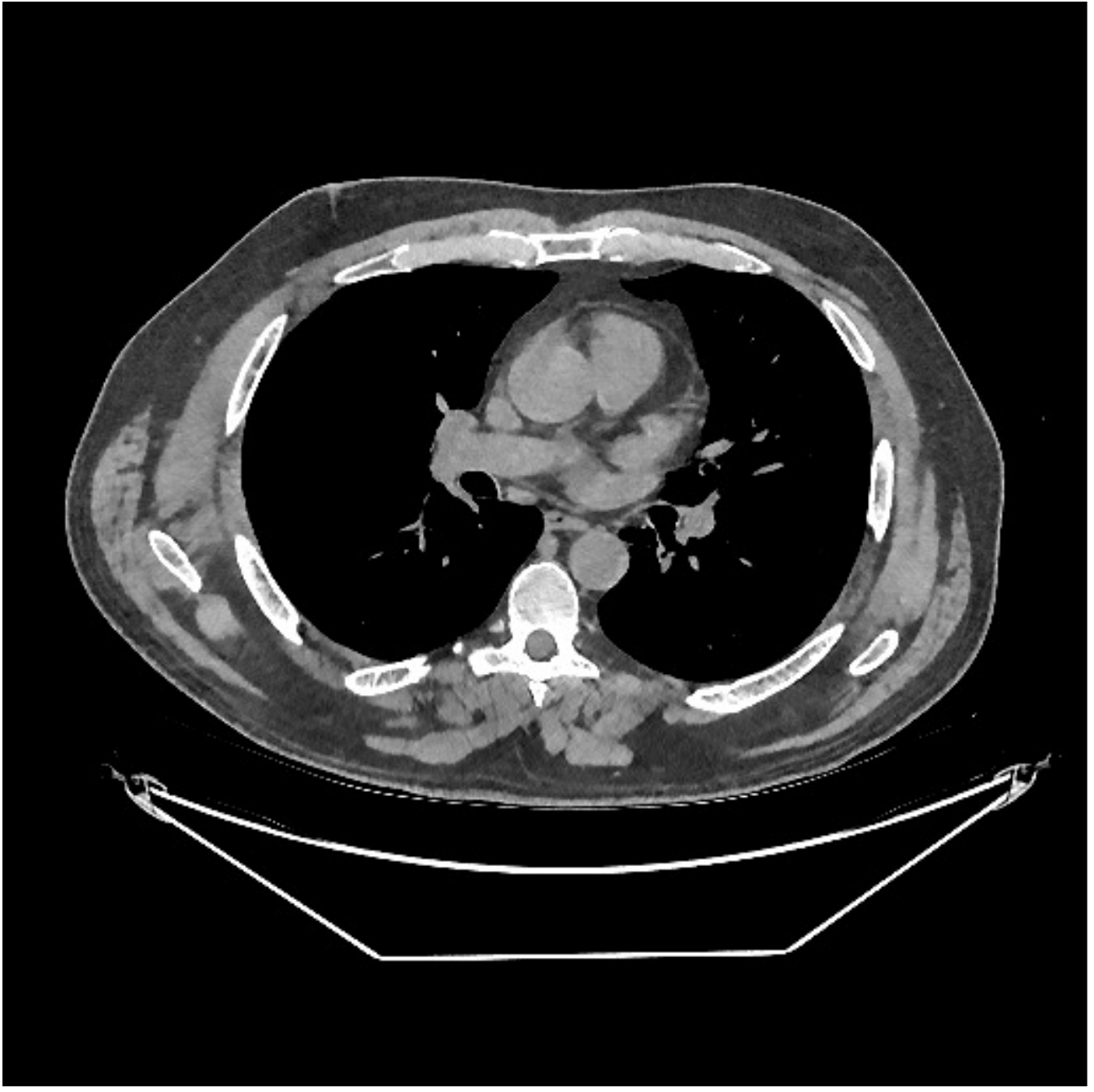}
		
	\end{tabular}

	\caption{Comparison of three reconstructed clinical images from different reconstruction methods for low-dose CT (images are magnified to better show differences; display window $[800, 1200]$~HU).}
	\label{fig:recon:ge}
\end{figure*}

For all the testing phantom and clinical images, the proposed BCD-Nets significantly improve the low-dose CT reconstruction accuracy, compared to the  conventional MBIR method 
using EP and/or the state-of-the-art MBIR method using $\ell_2$ prior with  a learned transform \cite{Zheng&etal:19TCI}. 
For all the testing phantom images, BCD-Net consistently achieves better reconstruction quality than ADMM-Net.
See Table~\ref{tab:rmse:xcat}, Figs.~\ref{fig:recon:ge} \& S.2, and Section~S.3.
In particular, BCD-Net accomplishes the both benefits of EP and image denoising (see Fig.~S.2);
and increasing the number of filters and thresholding parameters improves its reconstruction performance (see Table~\ref{tab:rmse:xcat}).

\begin{table}[!]	
	\vspace{-1.25pc}
	\centering
	\caption{RMSE (HU) of three reconstructed XCAT phantom images with different MBIR methods for low-dose CT$^\dagger$ ($\rho_0 = 10^4$ incident photons)}
	\label{tab:rmse:xcat}	 	
	\begin{tabular}{C{1.2cm}C{1cm}C{2.4cm}C{2.6cm}C{1.9cm}C{2.4cm}}			
		\toprule
		&  \small{EP}           
		& \specialcell[c]{\small Learned trans. \\ \small ($K \!=\! R \!=\! 8^2$) \cite{Zheng&etal:19TCI}}    
		& \specialcell[c]{\small ADMM-Net \\ \small ($K \!=\! R \!=\! 8^2$) \cite{Yang&etal:16NIPS}}  
		& \specialcell[c]{\small BCD-Net \\ \small ($K \!=\! R \!=\! 8^2$)}  
		& \specialcell[c]{\small \textbf{BCD-Net} \\ \small ($K \!=\! 10^2, R \!=\! 8^2$)}    
		\\
		\midrule
		\small{Test $\#1$}                   & 39.4       &  36.5     &31.6   & 30.7     &\textbf{27.5}    \\
		\small{Test $\#2$}                   & 39.6       &  37.8     &32.0   & 31.4     & \textbf{29.2}    \\
		\small{Test $\#3$}                   & 37.1       &  34.0     &32.0   & 30.6     & \textbf{27.7}    \\
		\bottomrule 
		\multicolumn{6}{l}{$^\dagger$See reconstructed images and error images in Fig.~S.2 and  Fig.~S.3, respectively.}
	\end{tabular}
	\vspace{-1.0pc}
\end{table}

\subsubsection{Generalization Capability Comparisons.}

The proposed BCD-Net has significantly better generalization capability than a state-of-the-art (non-iterative) deep regression NN, FBPConvNet \cite{Jin&etal:17TIP}.
Clinical scan experiments in Fig.~\ref{fig:recon:ge} indicate that deep regression NNs, e.g., FBPConvNet, can have high overfitting risks, 
while our proposed BCD-Net has low overfitting risks, and gives more stable reconstruction.
These show that MBIR modules benefit regularizing overfitting artifacts of regression NNs.

The BCD-Net result in the second row of Fig.~\ref{fig:recon:ge} shows non-uniform spatial resolution or noise;
see blurry artifacts particularly around the center of the reconstructed image.
One can reduce such blurs by including the technique of controlling local spatial resolution or noise in the reconstructed images \cite{Fessler&Rogers:96TIP} to MBIR modules.

\section{Conclusions}
\label{sec:conclusions}

The proposed BCD-Net uses layer-wise autoencoding CNNs and achieves significantly more accurate low-dose CT reconstruction, 
compared to the state-of-the-art MBIR method using a learned transform \cite{Zheng&etal:19TCI}.
BCD-Net provides better reconstruction quality, compared to a state-of-the-art iterative NN, ADMM-Net \cite{Yang&etal:16NIPS}.
Taking both benefits of MBIR and low-complexity CNN (i.e., convolutional autoencoder), BCD-Net significantly improves the generalization capability, compared to a state-of-the-art (non-iterative) deep regression NN, FBPConvNet \cite{Jin&etal:17TIP}.
In addition, applying faster numerical solvers, e.g., APG-M, to MBIR modules leads to faster and more accurate BCD-Net, and those with sufficient iterations can lead to the sequence convergence.
Future work will explore BCD-Net with local spatial resolution controls \cite{Fessler&Rogers:96TIP}, to reduce blur around the center of reconstructed images.

\subsubsection{Acknowledgments}
This work is supported in part by NSFC 61501292 and NIH U01 EB018753.
The authors thank GE Healthcare for supplying the clinical data. 
The authors thank Zhipeng Li for his help with debugging the codes.

 \bibliographystyle{splncs04.bst}
 \bibliography{referenceBibs_Bobby.bib}

\begin{thebibliography}{10}
\providecommand{\url}[1]{\texttt{#1}}
\providecommand{\urlprefix}{URL }
\providecommand{\doi}[1]{https://doi.org/#1}

\bibitem{Aggarwal&Mani&Jacobs:18TMI}
Aggarwal, H.K., Mani, M.P., Jacob, M.: {MoDL}: Model based deep learning
  architecture for inverse problems. IEEE Trans. Med. Imag.  \textbf{38}(2),
  {394--405} (Feb 2019)

\bibitem{Beck&Teboulle:09SIAM}
Beck, A., Teboulle, M.: A fast iterative shrinkage-thresholding algorithm for
  linear inverse problems. SIAM J. Imaging Sci.  \textbf{2}(1),  183--202 (Mar
  2009)

\bibitem{chan:17:pap}
Chan, S.H., Wang, X., Elgendy, O.A.: Plug-and-play {ADMM} for image
  restoration: fixed-point convergence and applications. IEEE Trans. Comput.
  Imag.  \textbf{3}(1),  {84--98} (Mar 2017)

\bibitem{Chun&Fessler:18IVMSP}
Chun, I.Y., Fessler, J.A.: Deep {BCD}-net using identical encoding-decoding
  {CNN} structures for iterative image recovery. In: Proc. IEEE IVMSP Workshop.
  Zagori, Greece (Jun 2018)

\bibitem{Chun&Fessler:18arXiv}
Chun, I.Y., Fessler, J.A.: Convolutional analysis operator learning:
  {A}cceleration and convergence. \emph{submitted}  (Jan 2019),
  \url{https://arxiv.org/abs/1802.05584}

\bibitem{Chun&etal:18arXiv:momnet}
Chun, I.Y., Huang, Z., Lim, H., Fessler, J.A.: {Momentum-Net}: {F}ast and
  convergent recurrent neural network for inverse problems. \emph{preprint}
  (Feb 2019)

\bibitem{Fessler&Rogers:96TIP}
Fessler, J.A., Rogers, W.L.: Spatial resolution properties of
  penalized-likelihood image reconstruction methods: {Space-invariant}
  tomographs. IEEE Trans. Image Process.  \textbf{5}(9),  1346--58 (Sep 1996)

\bibitem{Jin&etal:17TIP}
Jin, K.H., McCann, M.T., Froustey, E., Unser, M.: Deep convolutional neural
  network for inverse problems in imaging. IEEE Trans. Image Process.
  \textbf{26}(9),  4509--4522 (Sep 2017)

\bibitem{Kingma&Ba:15ICLR}
Kingma, D.P., Ba, J.L.: Adam: {A} method for stochastic optimization. In: Proc.
  ICLR $2015$. pp. 1--15. San Diego, CA (May 2015)

\bibitem{Rockafellar:76SIAMCO}
Rockafellar, R.T.: Monotone operators and the proximal point algorithm. SIAM J.
  Control Optm.  \textbf{14}(5),  877--898 (Aug 1976)

\bibitem{Romano&Elad&Milanfar:17SJIS}
Romano, Y., Elad, M., Milanfar, P.: The little engine that could:
  Regularization by denoising ({RED}). SIAM J. Imaging Sci.  \textbf{10}(4),
  1804--1844 (Oct 2017)

\bibitem{Segars&etal:08MP}
Segars, W.P., Mahesh, M., Beck, T.J., Frey, E.C., Tsui, B.M.: Realistic {CT}
  simulation using the {4D} {XCAT} phantom. Med. Phys.  \textbf{35}(8),
  3800--3808 (Jul 2008)

\bibitem{taylor:17:ewc-composite}
Taylor, A.B., Hendrickx, J.M., Glineur, F.: Exact worst-case performance of
  first-order methods for composite convex optimization. {SIAM J. Optim.}
  \textbf{27}(3),  {1283--1313} (Jan 2017)

\bibitem{Yang&etal:16NIPS}
Yang, Y., Sun, J., Li, H., Xu, Z.: Deep {ADMM-Net} for compressive sensing
  {MRI}. In: Proc. NIPS $29$. pp. 10--18. Long Beach, CA (Dec 2016)

\bibitem{Zheng&etal:19TCI}
Zheng, X., Chun, I.Y., Li, Z., Long, Y., Fessler, J.A.: Sparse-view {X}-ray
  {CT} reconstruction using $\ell_1$ prior with learned transform.
  \emph{submitted}  (Feb 2019), \url{http://arxiv.org/abs/1711.00905}

\end{thebibliography}



\section*{Supplementary material}

41# BCD-Net for Low-dose CT Reconstruction: Acceleration, Convergence, and Generalization – Supplementary Material

Il Yong Chun$^*$, Xuehang Zheng$^*$, Yong Long, and Jeffrey A. FesslerThis supplement shows additional experimental results to accompany the main manuscript. We use the prefix "S" for the numbers in sections, equations, figures, and tables in the supplementary material.

## S.1. Empirical convergence behaviors of BCD-Net

Fig. 1(a) shows that the relative differences between the parameters of the denoising CNNs $\mathcal{D}_{\boldsymbol{\theta}^{(l+1)}}$ in (1) at the previous and current layers, i.e., $\boldsymbol{\theta}^{(l-1)}$ and $\boldsymbol{\theta}^{(l)}$, tend to go zero as $l \to \infty$. This suggests that the trained convolutional autoencoders $\{\mathcal{D}_{\boldsymbol{\theta}^{(l+1)}} : \forall l \geq 0\}$ in (1) tend to converge as $l \to \infty$. Fig. 1(b) shows that an empirical Lipchitz constant of $\mathcal{D}_{\boldsymbol{\theta}^{(l)}}$, i.e., $\kappa^{(l)}$ in the bound $\|\mathcal{D}_{\boldsymbol{\theta}^{(l)}}(\mathbf{u}) - \mathcal{D}_{\boldsymbol{\theta}^{(l)}}(\mathbf{v})\|_2 \leq \kappa^{(l)} \|\mathbf{u} - \mathbf{v}\|_2$, $\forall \mathbf{u}, \mathbf{v}$, converges some constant less than 1. This suggests that the trained convolutional autoencoders $\{\mathcal{D}_{\boldsymbol{\theta}^{(l+1)}} : \forall l \geq 0\}$ in (1) become nonexpansive as $l \to \infty$.

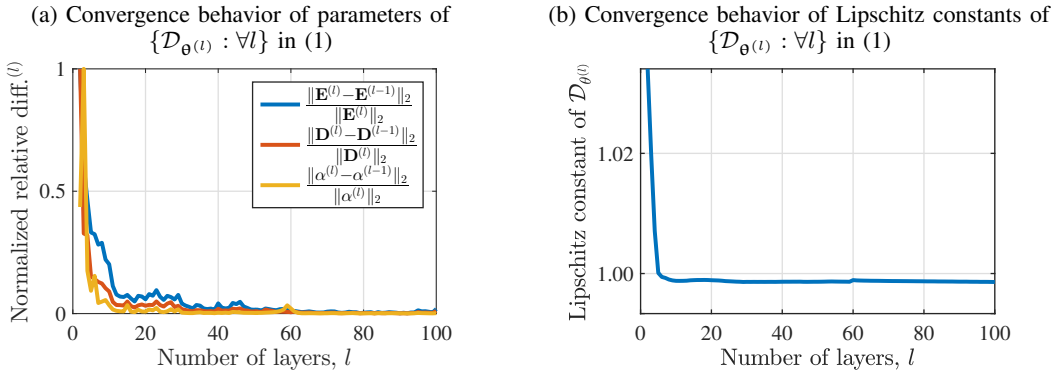

Fig. S.1. Empirical measures associated with the sequence convergence of BCD-Net (BCD-Net using the convolutional autoencoders in (1) with $\beta = 4 \times 10^6$ and $K = R = 8^2$, and 20 APG-M iterations for MBIR modules). (b) We calculated Lipschitz constants of the denoising CNNs in (1) by using all the combination pairs of ten training images.

## S.2. Parameter details for the reconstruction methods in Section 3.1

This section describes the parameter details of the reconstruction mentioned in Section 3.1:
- For the traditional filtered back projection (FBP) method, we used the Hanning window.
- For the three representative XCAT phantom images, the selected parameters are as follows. For EP MBIR, we chose the regularization parameter as $2^{16}$, and ran the relaxed linearized augmented Lagrangian method with ordered-subsets (relaxed OS-LALM) [1] for 50 iterations with 24 ordered subsets. For the MBIR method using a learned transform [2], we chose the regularization parameter as $2 \times 10^5$ and the hard-thresholding parameter as 20. Initialized with EP reconstructions, we used 2 iterations of relaxed OS-LALM with 4 ordered subsets for image updates, and ran for 1000 iterations. For both the BCD-Nets (initialized with FBP reconstructions), we set the regularization parameter $\beta$ as $4 \times 10^6$.
- For the two representative clinical images in the first and second rows of Fig. 2, the selected parameters are as follows. For EP MBIR, we chose the regularization parameter as $2^{16.5}$, and ran relaxed OS-LALM for 50 iterations with 12 ordered subsets. For BCD-Net (initialized with FBP reconstructions), we set the regularization parameter $\beta$ as $3 \times 10^6$.
- For the representative clinical image in third row of Fig. 2, the selected parameters are as follows. For EP MBIR, we chose the regularization parameter as $2^2$, and ran relaxed OS-LALM for 50 iterations with 6 ordered subsets. For BCD-Net (initialized with FBP reconstruction), we set the the regularization parameter $\beta$ as $0.08$.

## S.3. Additional experimental results

This section includes reconstructed XCAT phantom images and corresponding error images that give the results in Table 1, and some additional reference images for all reconstruction experiments described in the main paper.

Supplementary material dated July 25, 2019.
The authors indicated by asterisks ($^*$) equally contributed to this work.
Il Yong Chun and Jeffrey A. Fessler are with the Department of Electrical Engineering and Computer Science, The University of Michigan, Ann Arbor, MI 48019 USA (email: {iychun, fessler}@umich.edu).
Xuehang Zheng and Yong Long are with the University of Michigan - Shanghai Jiao Tong University Joint Institute, Shanghai Jiao Tong University, Shanghai 200240, China (email: {zhxhang, yong.long}@sjtu.edu.cn).

Fig. S.2 shows the three reconstructed XCAT phantom images with different MBIR methods, that corresponds to the results in Table 1. For all testing phantom images, the proposed BCD-Nets significantly improve the low-dose CT reconstruction accuracy, compared to the state-of-the-art MBIR method using a learned transform [2] and the conventional MBIR method using EP. Compared to ADMM-Net, BCD-Net better reconstructs bone structures in general. Compare results in Fig. S.2(c) and Fig. S.2(d). In particular, reconstructed images via ADMM-Net have some severe artifacts around the bone regions; see zoom-ins in Fig. S.2(c). We conjecture that additional dual variable updates incorporated in MBIR modules of ADMM-Net affect its MBIR accuracy.

Fig. S.3 shows the error images that correspond to the reconstructed images in Fig. S.2. In particular for test samples #2 and #3, reconstructed images via ADMM-Net have rougher error maps compared to those via BCD-Net. Compare results in Fig. S.3(c) and Fig. S.3(d).

Fig. S.4 shows the low-dose FBP reconstructions, and additionally provides reference images, i.e., ground truth or high-dose FBP reconstructions.

## REFERENCES

[1] Nien, H., Fessler, J.A.: Relaxed linearized algorithms for faster X-ray CT image reconstruction. IEEE Trans. Med. Imag. **35**(4), 1090–1098 (Apr 2016)
[2] Zheng, X., Chun, I.Y., Li, Z., Long, Y., Fessler, J.A.: Sparse-view X-ray CT reconstruction using $\ell_1$ prior with learned transform. *submitted* (Feb 2019), http://arxiv.org/abs/1711.00905

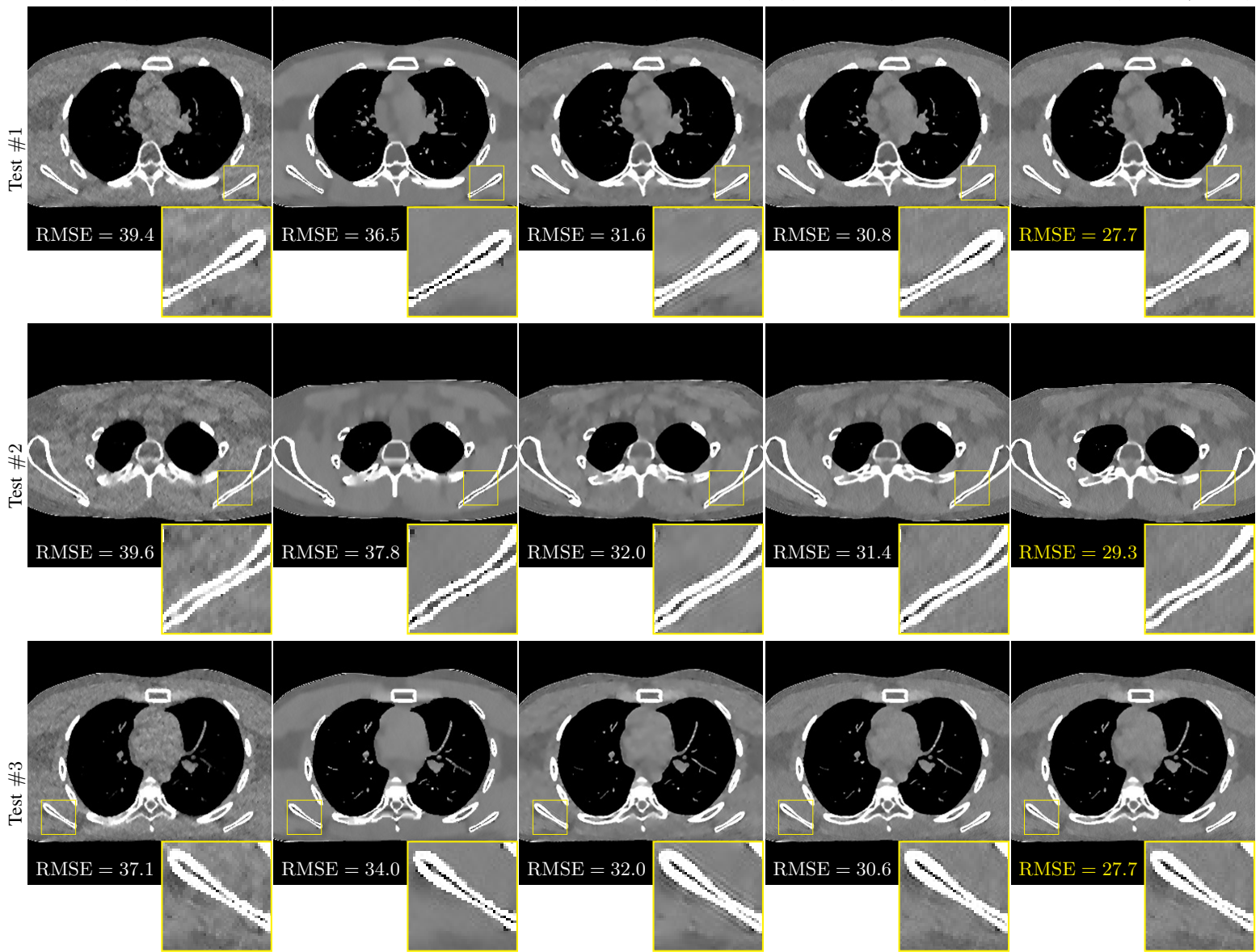

Fig. S.2. Comparison of three reconstructed XCAT phantom images with different MBIR methods (images are magnified to better show differences; display window is $[800, 1200]$ HU; and RMSE values are in HU).

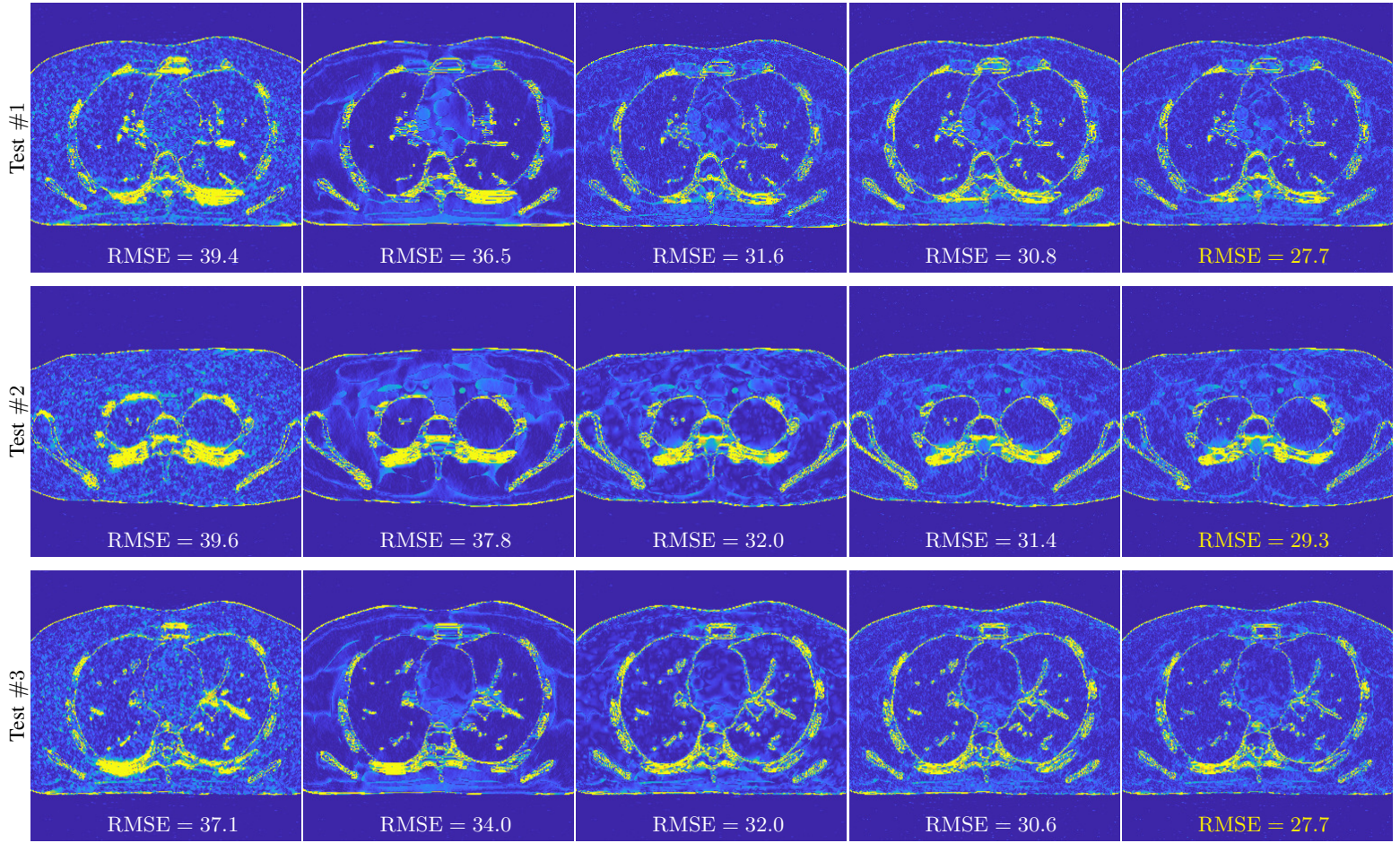

Fig. S.3. Error images of three reconstructed XCAT phantom images with different MBIR methods (images are magnified to better show differences; display window is $[0, 100]$ HU; and RMSE values are in HU).

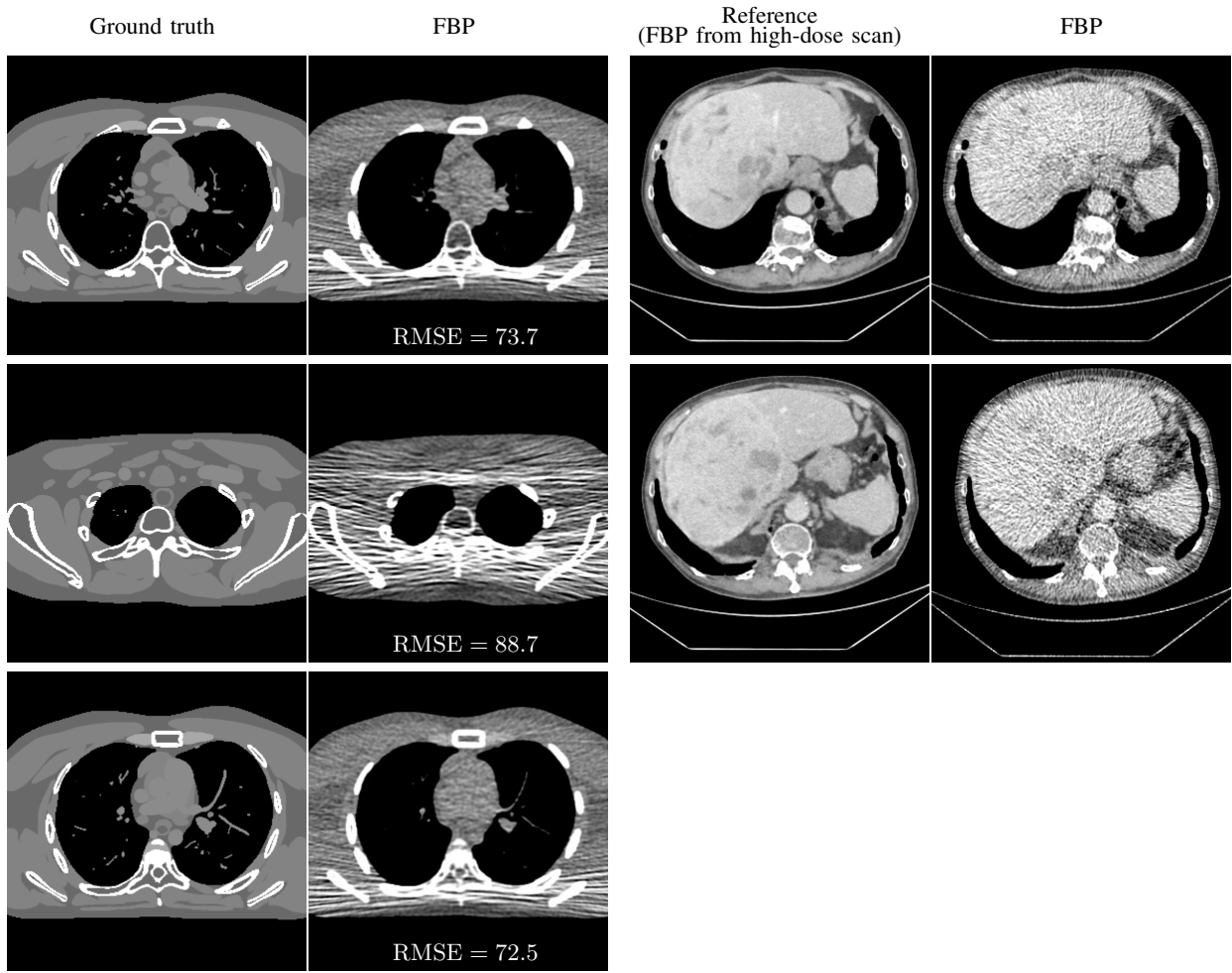

Fig. S.4. Reference images for the low-dose CT reconstruction experiments described in the main manuscript. The first and second columns show ground truth and FBP reconstruction of three slices of the XCAT phantom, respectively. The third and fourth columns show FBP reconstructions from high-dose ($\rho_0 = 5 \times 10^5$ incident photons) and low-dose ($\rho_0 = 1 \times 10^4$) scans of two representative clinical images, respectively. (Images are magnified to better show differences; display window is $[800, 1200]$ HU; and RMSE values are in HU.)



\end{document}